\documentclass[aps,pra,twocolumn,amsmath,amssymb,longbibliography]{revtex4-1}

\usepackage{graphicx}
\usepackage{dcolumn}
\usepackage{bm}
\usepackage{epsfig}
\usepackage{subfig}
\usepackage[all]{xy}
\usepackage{amsmath}
\usepackage{amsfonts}

\newcommand{\ba}{\begin{eqnarray}}
\newcommand{\ea}{\end{eqnarray}}
\newcommand{\be}{\begin{equation}}
\newcommand{\ee}{\end{equation}}
\newcommand{\bea}{\begin{eqnarray}}
\newcommand{\eea}{\end{eqnarray}}

\usepackage{theorem}

\theorembodyfont{\rmfamily}

\theoremstyle{break}

\def\QED{~\rule[-1pt]{5pt}{5pt}\par\medskip}

\begin{document}


\title{Maximal quantum Fisher information matrix}
\author{Yu Chen}
\affiliation{Department of Mechanical and Automation Engineering, The Chinese
University of Hong Kong, Shatin, Hong Kong}

\author{Haidong Yuan}
\email{hdyuan@mae.cuhk.edu.hk}
\affiliation{Department of Mechanical and Automation Engineering, The Chinese
University of Hong Kong, Shatin, Hong Kong}


\begin{abstract}
 We study the existence of the maximal quantum Fisher information matrix in multi-parameter quantum estimation, which bounds the ultimate precision limit. We show that when the maximal quantum Fisher information matrix exists, it can be directly obtained from the underlying dynamics. Examples are then provided to demonstrate the usefulness of the maximal quantum Fisher information matrix by deriving various tradeoff relations in multi-parameter quantum estimation and obtaining the bounds for the scalings of the precision limit.
\end{abstract}
\maketitle
\section{Introduction}
Quantum metrology, which uses quantum mechanical effects to improve the precision limit for parameter estimation, has attracted a lot attention recently\cite{Giovannetti2011,GIOV04,GIOV06,Fujiwara2008,Escher2011,Tsang2013,Rafal2012,Knysh2014,Jan2013,Rafal2014,Alipour2014,Chin2012,Tsang2011,Berry2013,Berry2015,pang2014, Liu2014,Yuan2015, YuanTime}. In quantum metrology, the precision limit is usually calibrated by the quantum Cram\'er-Rao bound $Cov(\hat{x})\geq J^{-1}$ where $Cov(\hat{x})$ denoting the covariance matrix of the estimator and $J$ denotes the quantum Fisher information matrix\cite{HELS67,HOLE82,Fisher,CRAM46,Rao}. While the quantum Cram\'er-Rao bound for single-parameter quantum estimation has been studied extensively, much less is known for multi-parameter quantum estimation, which differs from single-parameter quantum estimation in various ways\cite{Genoni2013,humphreys2013,vidrighin2014,crowley2014,Yue2014,Zhang2014,Gao2014,Tsang2014,Kok2015,Spagnolo2012,Datta,Yuan2016}. For single-parameter estimation, there always exists an optimal probe state that has the largest quantum Fisher information\cite{Yuan2015bridge}. For multi-parameter estimation, this is no longer the case, in general there is no single probe state that has the quantum Fisher information matrix(QFIM) dominating the QFIM of all other states. There could be two different states such that the corresponding QFIM $J_1$ and $J_2$ are not comparable, i.e., neither $J_1\geq J_2$ or $J_2\geq J_1$(here $J_1\geq J_2$ means $J_1-J_2$ is semi-positive definite)\cite{Manuel20042,Imai2007}. There is usually a corresponding optimal state for estimating each particular parameter, however these optimal states are typically incomparable unless one fixes a figure of merit that takes into account of the tradeoffs among the precision of different parameters. Such figure of merit is usually taken as $Tr[Cov(\hat{x})G]$ with $Cov(\hat{x})$ as the covariance matrix of the estimator and $G>0$ as a positive definite matrix.

An important question in multi-parameter quantum estimation is when there exists a global optimal state with a maximal QFIM, $J^{\max}$, such that $J^{\max}\geq J$ for all QFIM of other states, furthermore if such $J^{\max}$ exists how to compute it. Finding the maximal QFIM is useful in gauging the performance of practical estimation protocols as it provides a lower bound on the ultimate precision regardless of the choice of $G$ in the figure of merit since  $Tr[Cov(\hat{x})G]\geq Tr[J^{-1}G]\geq Tr[(J^{\max})^{-1}G]$ for all $G$. $J^{\max}$ thus sets a limit on the ultimate precision, which is particularly useful when proving the ultimate precision is bounded by certain scalings.


In this article we show that the maximal QFIM is closely related to the quantification of the distances between quantum channels and can be computed directly from the Kraus operators of the quantum channels. We also show how various tradeoff relations in multi-parameter quantum estimation can be obtained from the maximal QFIM, and provide a procedure to bound the scalings of the precision limit in multi-parameter quantum estimation.

\section{Maximal quantum Fisher information matrix}
In quantum estimation, typically to estimate some parameters $x=(x_1,x_2,\cdots,x_m)$ encoded in the dynamics $\phi_x$, one needs to first prepare a probe state $\rho_0$, let it evolve under the dynamics $\rho_0\xrightarrow{\phi_x} \rho_x$, then by performing Positive Operator Valued Measurements(POVM), $\{E_y\}$, on the output state $\rho_x$, one gets the measurement result $y$ with the probability $p(y|x)=Tr(E_y\rho_x)$. From the Cram\'{e}r-Rao bound in statistical theory\cite{HELS67,HOLE82,CRAM46,Rao}, the standard deviation for any unbiased estimator of $x$ is then bounded below by the Fisher information matrix: $nCov(\hat{x})\geq I^{-1}(x),$ where $n$ is the number of times that the procedure is repeated, $I(x)$ is the Fisher information matrix with the $ij$-th entry given by
$I_{ij}(x)=\int p(y|x)\frac{\partial lnp(y|x)}{\partial x_i}\frac{\partial lnp(y|x)}{\partial x_j} dy$\cite{Fisher}. The Fisher information matrix can be further bounded by the quantum Fisher information matrix(QFIM), which gives the quantum Cram\'er-Rao bound\cite{HELS67,HOLE82,BRAU94}
\begin{equation}
\label{eq:J}
nCov(\hat{x})\geq I^{-1}(x)\geq J^{-1}(\rho_x),
\end{equation}
where the $ij$-th entry of $J(\rho_x$) is given by $J_{ij}(\rho_x)=\frac{1}{2}Tr[\rho_x(L_iL_j+L_jL_i)]$, here the symmetric logarithmic derivative $L_i$ is the solution to the equation $\frac{\partial \rho_x}{\partial x_i}=\frac{1}{2}(\rho_xL_i+L_i\rho_x)$. In multi-parameter quantum estimation the inequality $I^{-1}(x)\geq J^{-1}(\rho_x)$ is usually not saturable due to the incompatibility of measurements for different parameters. One such tradeoff under separable measurement is manifested in the Gill-Massar(GM) inequality $Tr[J^{-1}(\rho_x)I(x)]\leq m-1$, here $m$ is the dimension of the system\cite{Gill2000}.

The precision of estimating $x$ from quantum states $\rho_x$ is also closely related to the Bures distance between $\rho_x$ and its infinitesimal neighboring states $\rho_{x+dx}$ \cite{HELS67,HOLE82,BRAU94}:
 \begin{equation}
\label{eq:BJ}
d^2_{Bures}(\rho_x,\rho_{x+dx})=\sum_{ij}\frac{1}{4}J_{ij}(\rho_x)dx_idx_j.
\end{equation}
Here the Bures distance $d_{Bures}$ is defined as
$d_{Bures}(\rho_1,\rho_2)=\sqrt{2-2F_B(\rho_1,\rho_2)},$
where $F_B(\rho_1,\rho_2)=\sqrt{\rho_1^{\frac{1}{2}}\rho_2\rho_1^{\frac{1}{2}}}$ is the fidelity, and $J_{ij}(\rho_x)$ denotes the $ij$-th entry of the QFIM. 

For a given quantum channel $K_x$, possibly aided with ancillary system, the output state is $\rho_x=K_x\otimes I_A(\rho_{SA})$(here $I_A$ denotes the identity operator on the ancillary system and $\rho_{SA}$ denotes a state of system+ancillary), then
\begin{eqnarray}
\label{eq:def}
\aligned
&d^2_{Bures}[K_x\otimes I_A(\rho_{SA}),K_{x+dx}\otimes I_A(\rho_{SA})]\\
=&\sum_{ij}\frac{1}{4}J_{ij}(\rho_x)dx_idx_j.
\endaligned
\end{eqnarray}

The existence of the maximal QFIM is then related to the distance between $K_x$ and $K_{x+dx}$, which corresponds to the maximal Bures distance between the output states over the input state $\rho_{SA}$. Specifically a maximal QFIM exists if and only if there exists a state $\rho^{\max}_{SA}$ achieving $\max_{\rho_{SA}}d^2_{Bures}[K_x\otimes I_A(\rho_{SA}),K_{x+dx}\otimes I_A(\rho_{SA})]$ for all $dx$(we will always assume $dx$ is taken sufficiently small so the second order expansion is valid). This can be seen in two steps: first if there exists a state $\rho^{\max}_{SA}$ achieves $\max_{\rho_{SA}}d^2_{Bures}[K_x\otimes I_A(\rho_{SA}),K_{x+dx}\otimes I_A(\rho_{SA})]$ for any $dx$, then the corresponding quantum Fisher information matrix, which we denote as $J^{\max}$, dominates all other QFIM, i.e., $J^{\max}\geq J(\rho_x)$ for all $\rho_x$. This is because for any probe state $\rho_{SA}$, and any $dx=(dx_1,dx_2,\cdots,dx_m)$, we have
\begin{eqnarray}
\aligned
&\sum_{ij}\frac{1}{4}J_{ij}(\rho_x)dx_idx_j\\
=&d^2_{Bures}(\rho_x,\rho_{x+dx})\\
=&d^2_{Bures}[K_x\otimes I_A(\rho_{SA}),K_{x+dx}\otimes I_A(\rho_{SA})]\\
\leq& \max_{\rho_{SA}}d^2_{Bures}[K_x\otimes I_A(\rho_{SA}),K_{x+dx}\otimes I_A(\rho_{SA})]\\
=&d^2_{Bures}[K_x\otimes I_A(\rho^{\max}_{SA}),K_{x+dx}\otimes I_A(\rho^{\max}_{SA})]\\
=&\sum_{ij}\frac{1}{4}J_{ij}^{\max}dx_idx_j,
\endaligned
\end{eqnarray}

 Thus $dx J^{\max} dx^T\geq dx J(\rho_x)dx^T$ for all $dx$, which implies $J^{\max}\geq J(\rho_x)$, i.e., $J^{\max}$ dominates all $J(\rho_x)$. Second if there does not exist a single probe state $\rho^{\max}_{SA}$ that achieves $\max_{\rho_{SA}}d^2_{Bures}[K_x\otimes I_A(\rho_{SA}),K_{x+dx}\otimes I_A(\rho_{SA})]$ for all $dx$, then there is no state that has QFIM dominating all $J(\rho_x)$. As if there exists a dominating maximal QFIM $J^{\max}$, which is the QFIM of some probe state $\tilde{\rho}_{SA}$, however since there does not exist a single probe state $\rho_{SA}$ that achieves $\max_{\rho_{SA}}d^2_{Bures}[K_x\otimes I_A(\rho_{SA}),K_{x+dx}\otimes I_A(\rho_{SA})]$ for all $dx$, thus there exists at least one $dx$ such that $d^2_{Bures}[K_x\otimes I_A(\tilde{\rho}_{SA}),K_{x+dx}\otimes I_A(\tilde{\rho}_{SA})]<d^2_{Bures}[K_x\otimes I_A(\rho^{op}_{SA}),K_{x+dx}\otimes I_A(\rho^{op}_{SA})]$, where $\rho^{op}_{SA}$ denotes the optimal state achieving the $\max_{\rho_{SA}}d^2_{Bures}[K_x\otimes I_A(\rho_{SA}),K_{x+dx}\otimes I_A(\rho_{SA}]$ for the chosen $dx$. Then
\begin{eqnarray}
\sum_{ij}\frac{1}{4}J_{ij}^{\max}dx_idx_j<\sum_{ij}\frac{1}{4}J^{op}_{ij}dx_idx_j,
\end{eqnarray}
i.e., $dx J^{\max} dx^T <dx J^{op} dx^T$, $J^{\max}$ is thus not a dominating QFIM for $J^{op}$.

We then show how to compute $J^{\max}$ directly from the quantum channels. We will first show the unitary case then extend to the noisy case. This follows the treatment in \cite{Yuan2015bridge}.   

We first show how to compute $J^{\max}$ from the unitary channel $U_x$. For any unitary $U$, we denote $e^{-iE^{U}_j}$ as the eigenvalues of $U$, where $E^{U}_j\in(-\pi,\pi]$ for $1\leq j\leq d$ (here $d$ denotes the dimension of $U$), which we call eigen-angles of $U$, and arrange $E^{U}_{\max}=E^{U}_1\geq E^{U}_2\geq \cdots \geq E^{U}_d=E^{U}_{\min}$ in decreasing order. Then
$\min_{\rho_0}F_B(\rho_0,U\rho_0 U^\dagger)=\cos\frac{E^{U}_{\max}-E^{U}_{\min}}{2}$ if $E^{U}_{\max}-E^{U}_{\min}\leq \pi$\cite{Fung2}. Denote $C(U)=\frac{E^{U}_{\max}-E^{U}_{\min}}{2}$, then $\min_{\rho_0}F_B(\rho_0,U\rho_0 U^\dagger)=\cos C(U)$. Since $E^{U\otimes I_A}_{\max}=E^{U}_{\max}$ and $E^{U\otimes I_A}_{\min}=E^{U}_{\min}$ we also have $\min_{\rho_{SA}}F_B(\rho_{SA},U\otimes I_A\rho_{SA} U^\dagger\otimes I_A)=\cos C(U)$.  
If the evolution is governed by $U_x=e^{-iH(x)T}$, with the aid of an ancillary system, $\rho_x=U_x\otimes I_A \rho_{SA}U_x^\dagger\otimes I_A$ and $\rho_{x+dx}=U_{x+dx}\otimes I_A \rho_{SA}U_{x+dx}^\dagger\otimes I_A$, it's then easy to see that $\max_{\rho_{SA}} d^2_{Bures}(\rho_x,\rho_{x+dx})=2-2\min_{\rho_{SA}}F_B(\rho_{SA},U'\otimes I_A\rho_{SA} U'^\dagger\otimes I_A)$
where $U'=U_x^\dagger U_{x+dx}$. From Eq.(\ref{eq:def}) we then get
\begin{eqnarray}
\label{eq:defJ}
\aligned
J^{\max}_{ij}dx_idx_j 
=8[1-\cos C(U_x^\dagger U_{x+dx})].
\endaligned
\end{eqnarray}
If the dynamics is continuous, then when $dx\rightarrow 0$, $U_x^\dagger U_{x+dx}\rightarrow I$, $C(U_x^\dagger U_{x+dx})\rightarrow 0$, thus up to the second order
\begin{eqnarray}
\label{eq:maxQFIphi2}
 J^{\max}_{ij}dx_idx_j=4C^2(U_x^\dagger U_{x+dx}).
  \end{eqnarray}
By comparing both sides of the equation we can then read out $J^{\max}$.

For a general quantum channel which maps from a $m_1$- to $m_2$-dimensional Hilbert space, the evolution can be represented by a Kraus operation $K_x(\rho^S)=\sum_{j=1}^d F_j(x)\rho^S F^\dagger_j(x)$, here the Kraus operators $F_j(x), 1\leq j\leq d$, are of the size $m_2\times m_1$, $\sum_{j=1}^d F^\dagger_j(x)F_j(x)=I_{m_1}$. It has been shown\cite{Yuan2015bridge}
that
\begin{eqnarray}
\label{eq:noisy}
\aligned
&\min_{\rho_{SA}} F_B[K_x\otimes I_A(\rho_{SA}), K_{x+dx}\otimes I_A(\rho_{SA})]\\
=& \max_{\|W\|\leq 1 }\frac{1}{2}\lambda_{\min}[K_W(x,x+dx)+K^\dagger_W(x,x+dx)],
\endaligned
\end{eqnarray}
 here $\lambda_{\min}[K_W(x,x+dx)+K^\dagger_W(x,x+dx)]$ denotes the minimum eigenvalue of $K_W(x,x+dx)+K^\dagger_W(x,x+dx)$ where $K_W(x,x+dx)=\sum_{ij}w_{ij}F^\dagger_i(x)F_j(x+dx)$ with $w_{ij}$ as the $ij$-th entry of a $d\times d$ matrix $W$ which satisfies $\|W\|\leq 1$($\|\cdot\|$ denotes the operator norm which equals to the maximum singular value). When $\max_{\|W\|\leq 1}\frac{1}{2}\lambda_{\min}[K_W(x,x+dx)+K^\dagger_W(x,x+dx)]$ can be analytically obtained we can read out the entries of $J^{\max}$ directly by comparing both sides of Eq.(\ref{eq:def}). When analytical solutions are not available, we can numerically obtain each entry of $J^{\max}$ by varying $dx$: for any given $dx$,  $\max_{\|W\|\leq 1}\frac{1}{2}\lambda_{\min}[K_W(x,x+dx)+K^\dagger_W(x,x+dx)]$ can be numerically computed via semi-definite programming as $\max_{\|W\|\leq 1}\frac{1}{2}\lambda_{\min}[K_W(x,x+dx)+K^\dagger_W(x,x+dx)]=$
\begin{eqnarray}
\label{eq:sdp}
\aligned
&max \qquad \frac{1}{2}t \\
s.t.\qquad &\left(\begin{array}{cc}
      I & W^\dagger  \\
      W & I \\
          \end{array}\right)\succeq 0,\\
      &    K_W(x,x+dx)+K^\dagger_W(x,x+dx)-tI \succeq 0.
          \endaligned
          \end{eqnarray}
To get the entries of $J^{\max}$, we can first choose $dx=(0,\cdots,0, dx_i,0,\cdots,0)$ and use Eq.(\ref{eq:def}) to get the diagonal entries $J^{\max}_{ii}$. We then choose $dx=(0,\cdots,dx_i,0,\cdots,0,dx_j,\cdots, 0)$ where only $dx_i$ and $dx_j$ are non-zero, Eq.(\ref{eq:def}) then gives the value of $1/4(J^{\max}_{ii}dx_i^2+2J^{\max}_{ij}dx_idx_j+J^{\max}_{jj}dx_j^2)$, the off-diagonal entries $J^{\max}_{ij}$ can then be obtained after subtracting the terms involving $J^{\max}_{ii}$ and $J^{\max}_{jj}$.

The semi-definite programming also has a dual form which can be used to identify the optimal probe state\cite{Yuan2015}. Specifically $\max_{\|W\|\leq 1}\frac{1}{2}\lambda_{\min}[K_W(x,x+dx)+K^\dagger_W(x,x+dx)]=\min_{\rho^S}\|M(\rho^S, K_x,K_{x+dx})\|_1$, here $\rho^S=Tr_A(\rho_{SA})$ with $\rho_{SA}=|\varphi_{SA}\rangle\langle\varphi_{SA}|$ and $M(\rho^S, K_x,K_{x+dx})$ is a $d\times d$ matrix with its $ij$-entry equal to $Tr[\rho^S F^\dagger_i(x)F_j(x+dx)]$, $\|\cdot\|_1$ denotes the trace norm which is the summation of singular values\cite{Yuan2015}. The dual semi-definite programming is given by $\min_{\rho^S}\|M(\rho^S, K_x,K_{x+dx})\|_1=$
\begin{eqnarray}
\label{eq:SDPrho}
\aligned
min \qquad &\frac{1}{2}Tr(P)+\frac{1}{2}Tr(Q) \\
s.t.\qquad &\left(\begin{array}{cc}
      P & M^\dagger(\rho^S, K_x,K_{x+dx})  \\
      M(\rho^S, K_x,K_{x+dx}) & Q \\
          \end{array}\right)\succeq 0,\\
       & \rho^S\succeq 0,
        Tr(\rho^S)=1,
          \endaligned
          \end{eqnarray}
here $P, Q$ are some Hermitian matrices. If the semi-definite programming outputs the same $\rho^S$ for different $dx$, then we have the global optimal probe state which has the QFIM attaining $J^{\max}$. 


\section{Applicaitons}
\subsection{Simultaneous estimation of the phase and dephasing strength}
Consider the dynamics $$K_x(\rho)=F_1(x)\rho F_1^\dagger(x)+F_2(x)\rho F_2^\dagger(x),$$ where $F_1(x)=\sqrt{\frac{1+\eta}{2}}\exp(-i\frac{\sigma_3}{2}\omega)$, $F_2(x)=\sqrt{\frac{1-\eta}{2}}\sigma_3\exp(-i\frac{\sigma_3}{2}\omega)$, here $\eta\in [0,1]$ denotes the dephasing noises and $\sigma_1=\left(\begin{array}{cc}
      0 & 1  \\
      1 & 0 \\
          \end{array}\right)$, $\sigma_2=\left(\begin{array}{cc}
      0 & -i  \\
      i & 0 \\
          \end{array}\right)$ and $\sigma_3=\left(\begin{array}{cc}
      1 & 0  \\
      0 & -1 \\
          \end{array}\right)$ are Pauli matrices. We would like to estimate $x=(\omega, \eta)$ simultaneously.
Let $\rho_{SA}$ be a pure input state for the extended channel $K_x\otimes I_A$, and $\rho^S=Tr_A(\rho_{SA})$. Then
$M(\rho^S, K_x,K_{x+dx})=\left(\begin{array}{cc}
      Tr[\rho^SF_1^\dagger(x)F_1(x+dx)] & Tr[\rho^SF_1^\dagger(x)F_2(x+dx)]  \\
      Tr[\rho^SF_2^\dagger(x)F_1(x+dx)] & Tr[\rho^SF_2^\dagger(x)F_2(x+dx)] \\
          \end{array}\right)$. By substituting the Kraus operators it is easy to show that, up to the second order, $$\|M(\rho^S, K_x,K_{x+dx})\|_1=1-\frac{1}{2}\rho^S_{11}\rho^S_{22}\eta^2d\omega^2-\rho_{11}\rho_{22}\frac{1}{2(1-\eta^2)}d\eta^2.$$ $\min\|M(\rho^S, K_x,K_{x+dx})\|_1$ is then achieved with $\rho^S_{11}=\rho^S_{22}=\frac{1}{2}$, i.e., $\min_{\rho^S}\|M(\rho^S, K_x,K_{x+dx})\|_1=1-\frac{\eta^2}{8}d\omega^2-\frac{1}{8(1-\eta^2)}d\eta^2$.
         In this case the optimal probe state that attains the maximum QFIM is any pure state that satisfies $\rho^S_{11}=\rho^S_{22}=\frac{1}{2}$. The simplest choice that satisfies this condition is $|+\rangle=\frac{|0\rangle+|1\rangle}{\sqrt{2}}$, which is independent of $dx$. And in this case the ancillary system is also not necessary for achieving the maximal QFIM.
         From Eq.(\ref{eq:def}) we have
          \begin{eqnarray}
          \aligned
          &J^{\max}_{ij}dx_idx_j\\
          =&8(1-\min_{\rho_{SA}}F_B[K_x\otimes I_A(\rho_{SA}), K_{x+dx}\otimes I_A(\rho_{SA})])\\
          =&8(1-\min_{\rho^S}\|M(\rho^S, K_x,K_{x+dx})\|_1)\\
          =&\eta^2d\omega^2+\frac{1}{1-\eta^2}d\eta^2,
          \endaligned
          \end{eqnarray}
          by comparing the coefficients of both sides we then get
          \begin{equation}
          J^{\max}=\left(\begin{array}{cc}
      \eta^2 & 0  \\
      0 & \frac{1}{1-\eta^2} \\
          \end{array}\right).
          \end{equation}
This is consistent with previous study\cite{vidrighin2014} but here with a much simpler derivation. From the quantum Cram\'er-Rao bound we then have $nCov(\hat{x})\geq (J^{\max})^{-1}$, here
$n$ is the number of times that the procedure is repeated and
 $Cov(\hat{x})=\left(\begin{array}{cc}
      A & D  \\
      D & B \\
          \end{array}\right)$,
 with $A=E[(\hat{\omega}-\omega)^2]$, $D=E[(\hat{\omega}-\omega)(\hat{\eta}-\eta)]$ and $B=E[(\hat{\eta}-\eta)^2]$.

In practice separable measurements are much easier to implement than joint measurements. If we restrict the measurements to be separable measurements and denote $I(x)$ as the Fisher information matrix, then $nCov(\hat{x})\geq I^{-1}(x)$ which can also be written as $\frac{1}{n}Cov(\hat{x})^{-1}\leq I(x)$. From Gill-Massar inequality $Tr[(J^{\max})^{-1}I(x)]\leq m-1$\cite{Gill2000} we have
 \begin{eqnarray}
 Tr[(J^{\max})^{-1}\frac{1}{n}Cov(\hat{x})^{-1}]\leq Tr[(J^{\max})^{-1}I(x)]\leq m-1,
 \end{eqnarray}
 here $m$ is the dimension of the system, in this case $m=2$. Thus
 \begin{eqnarray}
 \label{eq:comp}
 \aligned
 &Tr[(J^{\max})^{-1}Cov(\hat{x})^{-1}]\\
 =&Tr[(J^{\max})^{-1}\frac{1}{AB-D^2}\left(\begin{array}{cc}
      B & -D  \\
      -D & A \\
          \end{array}\right)]\\
          =&\frac{B}{AB-D^2}\frac{1}{J^{\max}_{11}}+\frac{A}{AB-D^2}\frac{1}{J^{\max}_{22}}\leq n,
 \endaligned
 \end{eqnarray}
 where $J^{\max}_{11}=\eta^2$ and $J^{\max}_{22}=\frac{1}{1-\eta^2}$. As $\frac{B}{AB-D^2}\geq \frac{1}{A}=\frac{1}{Var(\hat{\omega})}$, $\frac{A}{AB-D^2}\geq \frac{1}{B}=\frac{1}{Var(\hat{\eta})}$, this then gives a tradeoff relation
\begin{eqnarray}
\aligned
\frac{1}{Var(\hat{\omega})J^{\max}_{11}}+\frac{1}{Var(\hat{\eta})J^{\max}_{22}}\leq n.
\endaligned
\end{eqnarray}
This is equivalent to the tradeoff relations obtained in \cite{vidrighin2014}, which is a weaker version of Eq.(\ref{eq:comp}). 

\subsection{Simultaneous estimation of two parameters with non-commutating generators}
In previous example both the magnetic field and the dephasing are along the $z-$direction, thus commuting with each other. We now consider another example with non-commutating generators as $U(x)=e^{-iH(x)T}$, where
$H(x)=x_1\sigma_1 +x_2\sigma_2$, here the parameters to be estimated are $\vec{x}=(x_1, x_2)$. Since
\begin{eqnarray}
\aligned
U'&=U^\dagger(x) U(x+dx)\\
&=e^{i H(x)T}e^{-i H(x+dx)T},\\
\endaligned
\end{eqnarray}
 this is a unitary that can be written in the form as $e^{ia(x,dx)[\hat{k}(x,dx)\cdot \hat{\sigma}]}$(in the Bloch sphere representation $\hat{k}(x,dx)$ corresponds to the rotating axis while $a(x,dx)$ corresponds to the magnitude of rotating speed). $a(x,dx)$ can be easily computed as $\cos a(x,dx)=\cos (\|x\|_2T)\cos(\|x+dx\|_2T)+\frac{x}{\|x\|_2}\cdot \frac{x+dx}{\|x+dx\|_2}\sin(\|x\|_2T)\sin(\|x+dx\|_2T)$, here $\|x\|_2=\sqrt{x_1^2+x_2^2}$. The two eigen-angles of $U'$ are $E_{\max}=a(x,dx)$ and $E_{\min}=-a(x,dx)$, we thus have $\min_{\rho_{SA}}F_B[U_x\otimes I_A\rho_{SA}U_x^\dagger\otimes I_A,U_{x+dx}\otimes I_A\rho_{SA}U_{x=dx}^\dagger\otimes I_A]=\cos a(x,dx)$ which can be saturated when $\rho_{SA}$ is taken as the maximally entangled state $\frac{1}{\sqrt{2}}(|00\rangle+|11\rangle)$. $J^{\max}$ thus exists and from Eq.(\ref{eq:defJ}) we have  $$\sum_{ij}J^{\max}_{ij}dx_idx_j=8[1-\cos a(x,dx)].$$
By expanding $\cos a(x,dx)$ to the second order and comparing the coefficients, $J^{\max}$ can be obtained as
$J^{\max}=\left(\begin{array}{cc}
      J_{11} & J_{12}  \\
      J_{21} & J_{22} \\
          \end{array}\right)$
with $$J_{11}=4(\frac{x_1^2T^2}{x_1^2+x_2^2}+\frac{x_2^2}{x_1^2+x_2^2}\frac{\sin^2(\sqrt{x_1^2+x_2^2}T)}{x_1^2+x_2^2}),$$ $$J_{12}=J_{21}=4\frac{x_1x_2}{x_1^2+x_2^2}(T^2-\frac{\sin^2(\sqrt{x_1^2+x_2^2}T)}{x_1^2+x_2^2}),$$  $$J_{22}=4(\frac{x_2^2T^2}{x_1^2+x_2^2}+\frac{x_1^2}{x_1^2+x_2^2}\frac{\sin^2(\sqrt{x_1^2+x_2^2}T)}{x_1^2+x_2^2}).$$
As $(J^{\max})^{-1}=\frac{1}{J_{11}J_{22}-J_{12}J_{21}}\left(\begin{array}{cc}
      J_{22} & -J_{12}  \\
      -J_{21} & J_{11} \\
          \end{array}\right)$, thus
          \begin{equation}
          \label{eq:cov2}
          nCov(\hat{x})\geq \frac{1}{J_{11}J_{22}-J^2_{12}}\left(\begin{array}{cc}
      J_{22} & -J_{12}  \\
      -J_{21} & J_{11} \\
          \end{array}\right).
           \end{equation}
           This holds for all probe states, all POVM and all unbiased estimators. It thus sets a lower bound on the ultimate precision of $x_1$ and $x_2$. Note that if we know $x_2$(or $x_1$) a-priori then the ultimate precision of estimating $x_1$(or $x_2$) alone is given by $nVar(\hat{x}_1)\geq \frac{1}{J_{11}}$(or $nVar(\hat{x}_2)\geq \frac{1}{J_{22}}$). However when both parameters are not known a-priori, we have $nVar(\hat{x}_1)\geq \frac{J_{22}}{J_{11}J_{22}-J^2_{12}}$ and $nVar(\hat{x}_2)\geq \frac{J_{11}}{J_{11}J_{22}-J^2_{12}}$. As the off-diagonal term $J_{12}$ is not zero, the minimum variance for each parameter is strictly larger than the single parameter case. The non-zero off-diagonal terms in $J^{\max}$ thus quantifies the correlations between different parameters. Intuitively the larger the term is the bigger of the effect that the uncertainty in one parameter affects the ultimate precision of estimating the other parameter.

Various tradeoff relations can be derived from Eq.(\ref{eq:cov2}). For example consider the diagonal entries we have $nVar(\hat{x}_1)\geq \frac{J_{22}}{J_{11}J_{22}-J^2_{12}}$ and $nVar(\hat{x}_2)\geq \frac{J_{11}}{J_{11}J_{22}-J^2_{12}}\geq \frac{1}{J_{22}}$, this then leads to an uncertainly type tradeoff relation $Var(\hat{x}_1)Var(\hat{x}_2)\geq \frac{1}{n^2}\frac{1}{J_{11}J_{22}-J^2_{12}}$. Another tradeoff relation can be obtained from the determinant $det[nCov(\hat{x})]\geq det[(J^{\max})^{-1}]$, as $det[Cov(\hat{x})]=Var(\hat{x}_1)Var(\hat{x}_2)-Cov^2(\hat{x}_1,\hat{x}_2)$(here $Cov^2(\hat{x}_1,\hat{x}_2)=E[(\hat{x}_1-x_1)(\hat{x}_2-x_2)]$) and $det[(J^{\max})^{-1}]=\frac{1}{J_{11}J_{22}-J^2_{12}}$, we then get a stronger tradeoff relation as $Var(\hat{x}_1)Var(\hat{x}_2)-Cov^2(\hat{x}_1,\hat{x}_2)\geq \frac{1}{n^2}\frac{1}{J_{11}J_{22}-J^2_{12}}$. We note that these inequalities quantify the tradeoffs among the precision limit of different parameters in multi-parameter quantum estimation, which are different from the standard uncertainty relations that quantify the tradeoffs between the standard deviations of non-commuting observables.

\subsection{Standard quantum limit for multi-parameter estimation under certain noisy dynamics}
In this example we show that even when one can not identify the maximal QFIM, one can still use the distance between quantum channels to bound the scalings of the precision limit. For noisy dynamics it has been shown that the standard quantum limit is the generic scaling for the precision limit of single parameter quantum estimation\cite{Fujiwara2008,Escher2011,Rafal2012,Rafal2014}, here we provide a procedure to generalize it to multi-parameter quantum estimation.

\begin{figure}
  \centering
  \includegraphics[width=0.7\linewidth]{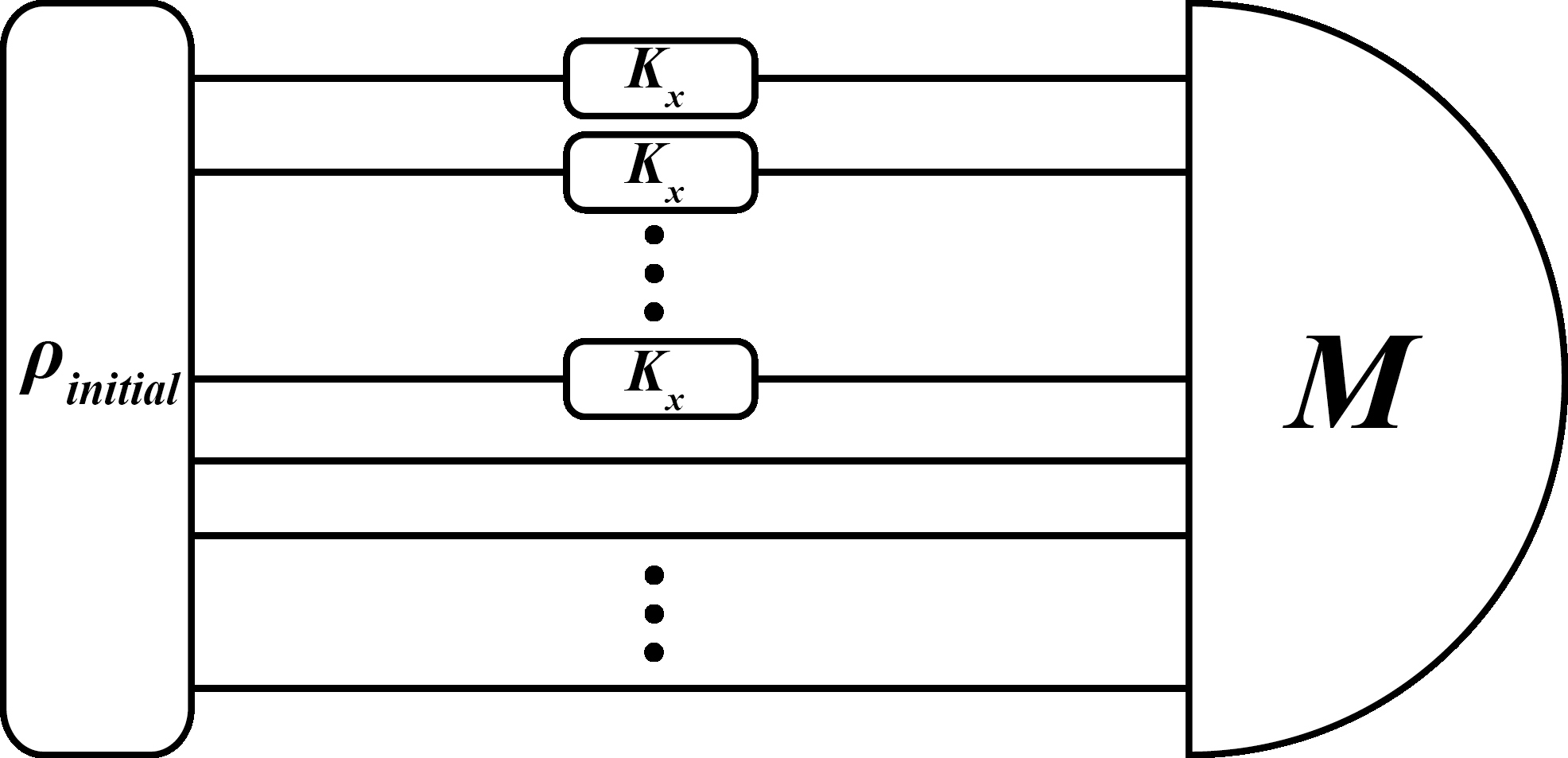}
  \caption{N quantum channels in parallel.}
  \label{fig:parallel}
\end{figure}

It is known\cite{Yuan2015bridge} that given any two channels  $K_1(\rho^S)=\sum_{j=1}^d F_{1j}\rho^S F^\dagger_{1j}$, $K_2(\rho^S)=\sum_{j=1}^d F_{2j}\rho^S F^\dagger_{2j}$,
\begin{eqnarray}
\label{eq:Nparallel}
\aligned
&2-2\min_{\rho_{SA}}F_B[K_1^{\otimes N}\otimes I_A(\rho_{SA}),K_2^{\otimes N}\otimes I_A(\rho_{SA})]\\
&\leq N\|2I-K_W-K_W^\dagger\|+N(N-1)\|I-K_W\|^2,
\endaligned
\end{eqnarray}
here $K^{\otimes N}$ denote $N$ channels in parallel, $K_W=\sum_{ij}w_{ij}F^\dagger_{1i}F_{2j}$, with $w_{ij}$ as the $ij$-th entry of a $d\times d$ matrix $W$ which satisfies $\|W\|\leq 1$. 
For N channels in parallel as in Fig.\ref{fig:parallel},
 where $K_x(\rho)=\sum_{i=1}^dF_i(x)\rho F_i^\dagger(x)$, we can substitute $K_1=K_x$ and $K_2=K_{x+dx}$ into Eq.(\ref{eq:Nparallel}) to get
 \begin{eqnarray}
\label{eq:Nparallelx}
\aligned
&2-2\min_{\rho_{SA}}F_B[K_x^{\otimes N}\otimes I_A(\rho_{SA}),K_{x+dx}^{\otimes N}\otimes I_A(\rho_{SA})]\\
&\leq N\|2I-K_W-K_W^\dagger\|+N(N-1)\|I-K_W\|^2,
\endaligned
\end{eqnarray}
 where $K_W=\sum_{ij}w_{ij}F^\dagger_{i}(x)F_{j}(x+dx)$. The observation is that if there exists a $W$ with $\|W\|\leq 1$ and a matrix $Q<\infty$ such that $\|I-K_W\|\leq \sum_{ij}Q_{ij}dx_idx_j$, then, as we will show, all QFIM $J(\rho_x)$ under the channel $K_x^{\otimes N}\otimes I_A$ scale at most linearly with $N$.

As in this case
\begin{eqnarray}
\aligned
&2-2\min_{\rho_{SA}}F_B[K_x^{\otimes N}\otimes I_A(\rho_{SA}),K_{x+dx}^{\otimes N}\otimes I_A(\rho_{SA})]\\
&\leq N\|2I-K_W-K_W^\dagger\|+N(N-1)\|I-K_W\|^2\\
&\leq N(\|I-K_W\|+\|I-K_W^\dagger\|)+N(N-1)\|I-K_W\|^2\\
&\leq 2N\sum_{ij}Q_{ij}dx_idx_j+O(dx^3).
\endaligned
\end{eqnarray}
Let $\rho_x=K_x^{\otimes N}\otimes I_A(\rho_{SA})$, and from Eq.(\ref{eq:def}) we have $\sum_{ij}J(\rho_x)_{ij}dx_idx_j=4\{2-2F_B[K_x^{\otimes N}\otimes I_A(\rho_{SA}),K_{x+dx}^{\otimes N}\otimes I_A(\rho_{SA})]\}$, thus
\begin{eqnarray}
\aligned
&\sum_{ij}J(\rho_x)_{ij}dx_idx_j\\
=&4\{2-2F_B[K_x^{\otimes N}\otimes I_A(\rho_{SA}),K_{x+dx}^{\otimes N}\otimes I_A(\rho_{SA})]\}\\
\leq &4\{2-2\min_{\rho_{SA}}F_B[K_x^{\otimes N}\otimes I_A(\rho_{SA}),K_{x+dx}^{\otimes N}\otimes I_A(\rho_{SA})]\}\\
\leq& 8N\sum_{ij}Q_{ij}dx_idx_j,
\endaligned
\end{eqnarray}
which implies $J(\rho_x)\leq 8NQ$ and this holds for any $\rho_x=K_x^{\otimes N}\otimes I_A(\rho_{SA})$. 
 The precision limit is then bounded by $Cov(\hat{x})\geq [nJ(\rho_x)]^{-1}\geq\frac{1}{8nN}Q^{-1}$, which scales as $\frac{1}{N}$, i.e., scales as the standard quantum limit. 

For example, consider the dephasing dynamics $$K_x(\rho)=F_1(x)\rho F_1^\dagger(x)+F_2(x)\rho F_2^\dagger(x),$$ where $F_1(x)=\sqrt{\frac{1+\eta}{2}}\exp(-i\frac{\sigma_3}{2}\omega)$, $F_2(x)=\sqrt{\frac{1-\eta}{2}}\sigma_3\exp(-i\frac{\sigma_3}{2}\omega)$ In this case there exists a $W=\begin{bmatrix}
      \cos(\xi d\omega)  & i\sin(\xi d\omega)  \\
      i\sin(\xi d\omega) & \cos(\xi d\omega) \\
          \end{bmatrix}$
 where $\xi=\frac{1}{2\sqrt{1-\eta^2}}$ for $\eta\neq 1$($\eta=1$ corresponds to the case of no dephasing error)\cite{Jan2013,Rafal2014}, such that up to the second order of $dx$
 \begin{eqnarray}
          \aligned
          &\|I-K_W\|\\
          =&\frac{1}{8(1-\eta^2)}\sqrt{\eta^2(d\eta^2+\eta d\omega^2)^2+4\eta^2d\eta^2d\omega^2}\\
          \leq& \frac{1}{8(1-\eta^2)}\sqrt{\eta^2(d\eta^2+\eta d\omega^2)^2+\eta(d\eta^2+\eta d\omega^2)^2}\\
          =& \frac{\sqrt{\eta^2+\eta}}{8(1-\eta^2)}(d\eta^2+\eta d\omega^2),
          \endaligned
          \end{eqnarray}
thus we have $\|I-K_W\|\leq \sum_{ij} Q_{ij}dx_idx_j$ here $x_1$ denote $\eta$, $x_2$ denote $\omega$ and $Q=\begin{bmatrix}
      \frac{\sqrt{\eta^2+\eta}}{8(1-\eta^2)}  & 0  \\
      0 & \frac{\eta\sqrt{\eta^2+\eta}}{8(1-\eta^2)}\\
          \end{bmatrix}$. Thus for any probe state $J(\rho_x)\leq 8NQ$ and the precision limit
$nCov(\hat{x})\geq J^{-1}\geq \frac{1}{8N}Q^{-1}=\frac{1}{N}\begin{bmatrix}
      \frac{1-\eta^2}{\sqrt{\eta^2+\eta}}  & 0  \\
      0 & \frac{1-\eta^2}{\eta\sqrt{\eta^2+\eta}}\\
          \end{bmatrix}$, which scales as $\frac{1}{N}$ for any $\eta\neq 1$.

\section{Conclusion}
We studied the maximal quantum Fisher information matrix for multi-parameter quantum estimation and showed its close relationship to the distance between quantum channels. The maximal quantum Fisher information matrix provides a lower bound on the precision limit regardless of the probe states and measurements, which can then be conveniently used to calibrate the practical protocols. Systematic procedure to obtain the maximal quantum Fisher information matrix is also provided and demonstrated with a few examples. We expect the connections established in this article can provide various insights for different scenarios of multi-parameter quantum estimation.

\end{document}